\documentclass[oneside,english]{amsart}
\usepackage[T1]{fontenc}
\usepackage[latin9]{inputenc}
\usepackage{mathrsfs}
\usepackage{amsthm}
\usepackage{amssymb}
\usepackage{graphicx}

\makeatletter
\numberwithin{equation}{section}
\numberwithin{figure}{section}
  \theoremstyle{remark}
  \newtheorem*{rem*}{\protect\remarkname}
  \theoremstyle{plain}
  \newtheorem*{prop*}{\protect\propositionname}

\makeatother

\usepackage{babel}
  \providecommand{\propositionname}{Proposition}
  \providecommand{\remarkname}{Remark}

\begin{document}

\title{Curvature in Hamiltonian Mechanics And The Einstein-Maxwell-Dilaton
Action}

\author{S. G. Rajeev}

\address{Department of Physics and Astronomy\\ Department of Mathematics\\
University of Rochester\ Rochester, NY 14627}

\email{s.g.rajeev@rochester.edu}

\date{Jan 15 2017}
\begin{abstract}
Riemannian geometry is a particular case of Hamiltonian mechanics:
the orbits of the hamiltonian $H=\frac{1}{2}g^{ij}p_{i}p_{j}$ are
the geodesics. Given a symplectic manifold $(\Gamma,\omega)$, a hamiltonian
$H:\Gamma\to\mathbb{R}$ and a Lagrangian sub-manifold $M\subset\Gamma$
we find a generalization of the notion of curvature. The particular
case $H=\frac{1}{2}g^{ij}\left[p_{i}-A_{i}\right]\left[p_{j}-A_{j}\right]+\phi$
of a particle moving in a gravitational, electromagnetic and scalar
fields is studied in more detail. The integral of the generalized
Ricci tensor w.r.t. the Boltzmann weight reduces to the action principle
$\int\left[R+\frac{1}{4}F_{ik}F_{jl}g^{kl}g^{ij}-g^{ij}\partial_{i}\phi\partial_{j}\phi\right]e^{-\phi}\sqrt{g}d^{n}q$
for the scalar, vector and tensor fields.
\end{abstract}

\maketitle

\section{Introduction }

 The theory of geodesics on a Riemannian manifold is a particular
case of Hamiltonian mechanics: they are simply the solutions of Hamilton's
equations for $H=\frac{1}{2}g^{ij}p_{i}p_{j}$. Is there a generalization
of Riemannian geometry corresponding to more general hamiltonians?
The three ideas we would like to generalize are those of distance,
volume and curvature. 

The volume is the easiest to generalize: the Boltzmann weight gives
a natural measure of integration in the phase space. Integrating out
the momenta gives the generalization of the Riemannian volume element
in configuration space (as long as $\int e^{-H}d^{n}p$ converges).

Recall that the ``reduced action'' (a.k.a ``eikonal'') $\sigma_{E}(Q,Q')=\int_{Q'}^{Q}p_{i}(E,q)dq^{i}$
of the trajectory of given energy $E$ connecting two points is a
good candidate for distance; although in the most general case it
is not positive or symmetric, let alone satisfy the triangle inequality.
If the Hamiltonian is an even function of momenta $H(q,p)=H(q,-p)$
we have time reversal invariance and the symmetry $\sigma_{E}(Q,Q')=\sigma_{E}(Q',Q)$
follows. If in addition $H(q,p)$ is a convex function of the momenta,
$\sigma_{E}(Q,Q')$ satisfies the triangle inequality. In particular,
for familiar mechanical systems it reduces to the Jacobi-Maupertuis
metric.

The most subtle notion to generalize is curvature. We will use the
second variation of the action to find a quantity $\mathcal{R}_{ij}(q,p)$
which transforms as a symmetric tensor under co-ordinate transformations
in configuration space and which reduces to the Riemann tensor in
Riemannian geometry: $\mathcal{R}_{ij}(q,p)=-g_{im}(q)p^{k}p^{l}R_{\ klj}^{m}(q)$
where two of the indices of the Riemann tensor are contracted by momentum.

An explicit formula for curvature in terms of derivatives of the hamiltonian
(up to fourth order) will be given. Its trace is the generalization
of the Ricci form; integrating over momentum gives the analogue of
the Ricci scalar density. We will show that in the case of most physical
interest, 

\[
H(q,p)=\frac{1}{2}g^{ij}(q)[p_{i}-A_{i}(q)][p_{j}-A_{j}(q)]+\phi(q)
\]

this Ricci scalar density is a natural unified action for gravity
coupled to electromagnetic and dilaton fields. Usually such actions
arise from much more complicated theories with many unwanted fields
(Kaluza-Klein, String theory etc.). 

To further strengthen our claim of having found the correct generalization,
we will show that in some simple but rather subtle cases (Lagrange
points, Penning trap) the positivity of curvature is sufficient for
stability; while negative curvature implies instability. 

The simple harmonic oscillator has constant positive curvature; it
also has finite diameter for the allowed subset of states of the configuration
space for  a given a energy. Its phase space has finite volume w.r.t.
to the Boltzmann measure. Thus the simple harmonic oscillator is the
mechanical analogue of the sphere $\mathbb{S}^{2}$. 

In Riemannian geometry, Myer's theorem says that the diameter is finite
when the curvature is bounded below by a positive constant. Under
the same condition, the Lichnerowicz theorem says that the spectrum
of the Laplacian has a gap (the first non-zero eigenvalue is bounded
below by a constant) . Perhaps there are generalizations to mechanical
systems. 

Let us spell out these ideas in some more detail before delving into
calculations.

\subsection{Distance}

Even in Riemannian geometry, the natural way to measure distance between
two points $Q',Q$ is to find the minimum over piece-wise differentiable
curves of the action

\[
S=\frac{1}{2}\int_{0}^{T}g_{ij}\dot{q}^{i}\dot{q}^{j}dt,\quad q(0)=Q',\quad q(T)=Q
\]

rather than the arc-length

\[
l=\frac{1}{2}\int_{0}^{T}\sqrt{g_{ij}\dot{q}^{i}\dot{q}^{j}}dt.
\]

The square root makes the arc-length tricky to differentiate as a
function of the curve . (Just as $|x|$ is not differentiable, unlike
$x^{2}$). In the jargon of high energy physics, the re-parametrization
invariance of $l$ is a ``gauge invariance'' that needs to be fixed;
$S$ is a ``gauge-fixed'' version of $l$. $S$ is called energy
in many mathematics textbooks\cite{doCarmoRiemGeom}, but the proper
mechanical analogue is action. 

Let $s_{T}(Q,Q')$ be the value of $S$ on the minimizing curve. It
is related to the Riemannian distance $d(Q,Q')$ (i.e., the minimum
of the arc-length) through 

\[
s_{T}(Q,Q')=\frac{d^{2}(Q,Q')}{2T}.
\]

It satisfies the Hamilton-Jacobi equation 

\[
\frac{1}{2}g^{ij}\partial_{Q^{i}}s_{T}\partial_{Q^{j}}s_{T}+\frac{\partial s_{T}}{\partial T}=0.
\]

Its Legendre transform

\[
\sigma_{E}(Q,Q')=\min_{T}\left[ET+s_{T}(Q,Q')\right],\quad\sigma_{E}(Q,Q')=\sqrt{2E}d(Q,Q')
\]

satisfies the stationary Hamilton-Jacobi equation (the eikonal equation
in the language of optics)

\[
\frac{1}{2}g^{ij}\partial_{Q^{i}}\sigma_{E}\partial_{Q^{j}}\sigma_{E}=E.
\]

Because $T,E$ only appear as overall factors in these formulas, it
is customary in geometry to choose units where they are set to constant
values ($T=\frac{1}{2}=E$) ; this corresponds to choosing a parametrization
where the velocity of the geodesic is unity.

For a more general mechanical system the action is still some integral
over the curve 

\[
S=\int L(q,\dot{q})dt
\]

but it may not be a purely quadratic function of the velocities. If
we drop the condition that $L(q,\dot{q})$ be quadratic in $\dot{q}$,
but insist that it is homogenous

\[
L(q,\lambda\dot{q})=\lambda^{r}L(q,\dot{q}),\quad\lambda,r>0
\]

we get the well-studied case of Finsler geometry\cite{ChernFinsler,RandersGRUnifiedEMPhysRev.59.195}.
As Chern points out, this is the case originally studied by Riemann
in his Habilitation; later, the name Riemannian geometry came to be
associated to the restricted case of a quadratic form. Finsler geometry
is so natural that Chern says\cite{ChernFinsler} 
\begin{quote}
``Finsler geometry is not a generalization of Riemannian geometry.
It is better described as Riemannian geometry without the quadratic
restriction''. 
\end{quote}
The typical hamiltonian of a mechanical system is a quadratic, but
not homogenous function of momenta 

\[
H(q,p)=\frac{1}{2}g^{ij}\left[p_{i}-A_{i}(q)\right]\left[p_{j}-A_{j}(q)\right]+\phi(q)
\]

corresponding to a Lagrangian

\[
L(q,\dot{q})=\frac{1}{2}g_{ij}\dot{q}^{i}\dot{q}^{j}+A_{i}\dot{q}^{i}-\phi(q).
\]

This describes a particle moving under the influence of gravitational,
electromagnetic and scalar forces. It is natural look for a geometry
associated to such (or even more general) hamiltonians. To paraphrase
Chern, we seek 
\begin{quote}
``Riemannian Geometry Without the Homogenous Restriction''.
\end{quote}
Such ideas go back to Hamilton\cite{Hamilton} himself, as noted by
Klimes\cite{Klimes}. 

\subsection{Hamiltonian Mechanics}

Let us recall some basic facts of mechanics. By Darboux's theorem\cite{WeinsteinLagrSubMflds}
every symplectic manifold $(\Gamma,\omega)$ can be covered by co-ordinate
charts such that in each chart the symplectic form has constant coefficients:

\[
\omega=dp_{i}dq^{i}.
\]

These co-ordinates satisfy the canonical Poisson brackets

\[
\left\{ q^{i},q^{j}\right\} =0=\left\{ p_{i},p_{j}\right\} ,\quad\left\{ p_{i},q^{j}\right\} =\delta_{i}^{j}
\]

A hamiltonian $H:\Gamma\to\mathbb{R}$ determines a family of curves
that pass through each point of $M$, satisfying Hamilton's equations 

\[
\dot{q}^{i}=H^{i},\quad\dot{p}_{i}=-H_{i}
\]

where 

\[
\dot{q}^{i}=\frac{dq^{i}}{dt},\quad H^{i}=\frac{\partial H}{\partial p_{i}},\quad H_{i}=\frac{\partial H}{\partial q^{i}}.
\]

These curves are the extrema of the action

\[
S=\int\left[p_{i}\dot{q}^{i}-H\right]dt.
\]

If we are also given a Lagrangian sub-manifold $M\subset\Gamma$,
locally $\Gamma$ can be identified\cite{WeinsteinLagrSubMflds} with
the co-tangent bundle $T^{*}M$. That is, there is a neighborhood
in which $M$ is determined by $p_{i}=0$ so that $q^{i}$ are co-ordinates
on it. An example is the case where $M$ is the configuration space
of a physical system.

Given a pair of points that are close enough and a time $T$, there
is a solution to Hamilton's equations with the boundary conditions 

\[
q^{i}(0)=Q^{i\prime},\quad q^{i}(T)=Q^{i}
\]

The action of this solution $s_{T}(Q,Q')$ has a Legendre transform

\[
\sigma_{E}(Q,Q')=\min_{T}\left[ET+s_{T}(Q,Q')\right]
\]

They satisfy the time dependent 

\[
H\left(Q,\partial_{Q}s_{T}\right)+\frac{\partial s_{T}}{\partial T}=0
\]

and stationary 

\[
H\left(Q,\partial_{Q}\sigma_{E}\right)=E
\]

versions of the Hamilton-Jacobi equation. Thus $\sigma_{E}(Q,Q')$
can be thought of as a generalization of the Riemannian distance function.
We will adopt terminology from optics and call $\sigma_{E}(Q,Q')$
the eikonal.

But there are important differences; it may not be a homogenous function
of $E$ unless the hamiltonian happens to have some sort of scale
symmetry. In general, it is not symmetric:

\[
\sigma_{E}(Q,Q')\neq\sigma_{E}(Q',Q)
\]

An example is a particle moving in a magnetic field (there is an explicit
calculation below). Thus, we will not be able to define a metric (in
the sense of topology) on the Lagrangian manifold $M$ (configuration
space) using $\sigma_{E}$. But that should not bother physicists
too much: we already gave that up when we allowed $g_{ij}$ to have
Lorentzian signature in relativistic mechanics.

\subsubsection{Time Reversal Invariant systems}

If the hamiltonian is time reversal invariant 
\[
H(q,p)=H(q,-p)
\]
 the eikonal will be a symmetric function 
\[
\sigma_{E}(Q,Q')=\sigma_{E}(Q',Q).
\]

An example is the case of a particle moving in a potential but with
no magnetic field:

\[
H(q,p)=\frac{1}{2}g^{ij}p_{i}p_{j}+V(q).
\]

In this example the stationary Hamilton-Jacobi equation can be rewritten
as 

\[
\frac{1}{2\left[E-V(Q)\right]}g^{ij}\partial_{Q^{i}}\sigma_{E}\partial_{Q^{j}}\sigma_{E}=1
\]

which is just the eikonal equation for the Jacobi-Maupertius metric 

\[
\hat{g}_{ij}(q)=[E-V(q)]g_{ij}(q).
\]

The trajectories must lie in the region where $V(q)<E$; they are
geodesics of the Jacobi-Maupertius metric on the manifold whose boundary
consist of turning points where $E=V(q)$. In particular, $\sigma_{E}(Q,Q')$
satisfies the triangle inequality and is a metric (in the sense of
topology) .

This example suggests that in the case of time reversal invariant
systems for which $H(q,p)$ is a convex function of momenta, (i.e.,
$H^{ij}(q,p)\equiv\frac{\partial^{2}H}{\partial p_{i}\partial p_{j}}$
is a positive matrix), the eikonal $\sigma_{E}(Q,Q')$ is a metric
in the sense of topology on some subset $M_{E}\subset M$ of allowed
configurations. It would be interesting to have a rigorous mathematical
proof of this.

\emph{Some Remarks:}
\begin{itemize}
\item Is there a version of Myer's theorem\cite{doCarmoRiemGeom}? That
is, given that the Ricci curvature(defined below) is bounded below
$\mathcal{R}\geq\omega^{2}>0$ , does it follow that $\sigma_{E}(Q,Q')\leq\pi\frac{E}{\omega}$?
We will see some elementary examples that suggest that this is true.
Is the fundamental group of $M_{E}$ finite? 
\item In the other direction, does $-\mathcal{R}\geq\omega^{2}>0$ and boundedness
of $\sigma_{E}$ imply that $M_{E}$ has infinite fundamental group?
This could have applications to ergodicity. 
\item When we pass to the quantum theory, $\sigma_{E}$ becomes the phase
of the wave function and the HJ equation becomes the Schrodinger equation.
In the toy model where $M$ is one-dimensional, it is possible to
find a quantum theory of gravity based on this interpretation\cite{HendersonRajeev}.
Perhaps it is of interest to see how much of that generalizes to higher
dimensions.
\end{itemize}

\subsection{Volume}

Given a hamiltonian $H:\Gamma\to\mathbb{R}$, there is a natural measure
of integration on phase space (motivated by Thermodynamics), the Boltzmann
weight(Of course, $n=\frac{1}{2}\dim\Gamma$.)

\[
d\mu_{H}=e^{-H}\frac{d^{n}pd^{n}q}{\left[2\pi\right]^{\frac{n}{2}}}.
\]

It is normalized to agree, in that special case, with the Riemannian
volume $\sqrt{g}d^{n}q$ on the configuration space (after integrating
out the momentum directions). Also, we have chosen units in which
the  temperature is equal to one. 

\subsection{Curvature}

As noted earlier, there is no obstruction to choosing local co-ordinates
in which the symplectic form has constant components. So, unlike a
Riemannian metric, a symplectic form does not uniquely determine a
connection or curvature. There are many torsion-less connections,
that preserve the symplectic form; there is no local obstruction to
choosing the curvature to be zero. There could be global obstructions
however. It is possible to choose a connection on a symplectic manifold
by a variational principle\cite{SymplecticConnections}. This is useful
in deformation quantization. None of this has any dependence on the
hamiltonian. 

Instead, we want to construct a curvature from the Hamiltonian that
measures the response of a mechanical system to small perturbations.
But a hamiltonian $H$, curvature. For example, there is a neighborhood
of every minimum of $H$ in which the Hamiltonian can be brought to
the Birkhoff Normal Form\cite{BambusiBirkhoffNormalForm}. Assuming
that the natural frequencies of small oscillations at the minimum
are not rationally related (which is the generic case) there is a
canonical transformation that brings $H$ to the quadratic form

\[
H(q,p)=\frac{1}{2}\sum_{k}\left[p_{k}^{2}+\omega_{k}^{2}q_{k}^{2}\right]+\cdots
\]

up to any desired order $\geq3$ in $p_{k},q_{k}$.

In Riemannian geometry, the infinitesimal deviation of geodesics (which
is determined by the second variation of the action) determines curvature.
It would be useful to have a generalization of curvature to more general
mechanical systems. For example, negative curvature could be an indication
of dynamical instability\cite{ArnoldCurvature,Govind3Body}.

Given a symplectic manifold $\Gamma$, a hamiltonian $H:\Gamma\to\mathbb{R}$
and a Lagrangian sub-manifold $M\subset\Gamma$ we will construct
a notion of curvature. The trick is to again consider the second variation
of the action. We will be able to write it as 

\[
\mathcal{S}_{1}=\int\left[\frac{1}{2}G_{ij}(q,p)\overset{\circ}{\xi}^{i}\overset{\circ}{\xi}^{j}-\frac{1}{2}\xi^{i}\xi^{j}\mathcal{R}_{ij}(q,p)\right]dt
\]

Here $\xi$ is the infinitesimal variation of the orbit, thought of
as a curve in $M$. Also, $\stackrel{\circ}{\xi}^{i}$ is a covariant
derivative of $\xi$ along the orbit. (The explicit formula is given
later). \emph{We do not attempt to define a covariant derivative (connection,
parallel transport etc.) along an arbitrary direction.}

$G_{ij}(q,p)$ is the inverse matrix of the second derivative of the
hamiltonian w.r.t. momentum 

\[
H^{ik}G_{kj}=\delta_{j}^{i},\quad H^{ij}=\frac{\partial H}{\partial p_{i}\partial p_{j}}.
\]

\emph{We will require that this second derivative $H^{ij}$ of the
hamiltonian be a positive matrix, so that the inverse exists}. (It
will be clear in most cases how to do a ``Wick Rotation'' to the
case (e.g., of Lorentzian signature) when $H^{ij}$ is only invertible
and not positive.) That is, we require that $H(q,p)$ is a convex
function of momenta. We will prove that the quantities $G_{ij}(q,p),\mathcal{R}_{ij}(q,p)$
transform as symmetric tensors under co-ordinate transformations $q^{i}\to\tilde{q}^{i}(q)$.
They are not in general homogenous functions of $p$. An explicit
expression for the curvature tensor $\mathcal{R}_{ij}(q,p)$ in terms
of derivatives (up to fourth order) of $H$ will be given. 

There is also an analogue of the Ricci tensor

\[
\mathcal{R}(q,p)=H^{ij}\mathcal{R}_{ij}(q,p)
\]

The generalization of the Ricci scalar-density is its average over
momentum:

\[
\mathfrak{R}(q)=\int\mathcal{R}(q,p)e^{-H(q,p)}\frac{d^{n}p}{\left[2\pi\right]^{\frac{n}{2}}}.
\]

\subsubsection{Riemannian Geometry}

In the particular case of Riemannian geometry

\[
H(q,p)=\frac{1}{2}g^{ij}(q)p_{i}p_{j}
\]

they reduce to the Riemann tensor $R_{\ klj}^{m}$, Ricci tensor $R_{ij}$
and the Ricci scalar-density as follows:

\[
\mathcal{R}_{ij}(q,p)=-g_{im}H^{k}H^{l}R_{\ klj}^{m}(q),\quad\mathcal{R}(q,p)=H^{k}H^{l}R_{kl}(q),\quad\mathfrak{R}(q)=\sqrt{g}R(q)
\]

where 

\[
H^{k}=g^{km}p_{m}.
\]

Thus $\mathscr{R}$ is the Einstein-Hilbert Lagrangian density for
GR (with Euclidean signature).

\subsubsection{Adding a Magnetic Field}

The hamiltonian 

\[
H=\frac{1}{2}g^{kl}\left[p_{k}-A_{k}\right]\left[p_{l}-A_{l}\right]
\]

leads to 

\[
\mathcal{R}_{ij}=-g_{im}H^{k}H^{l}R_{\ klj}^{m}+\frac{1}{4}F_{ik}F_{jl}g^{kl}+\frac{1}{2}H^{k}\left\{ \partial_{j}F_{ki}+\partial_{i}F_{kj}\right\} 
\]

\[
\mathcal{R}=H^{k}H^{l}R_{kl}+\frac{1}{4}F_{ik}F_{jl}g^{kl}g^{ij}+H^{k}g^{ij}\partial_{i}F_{kj}
\]

\[
\mathscr{R}=\left[R+\frac{1}{4}F_{ik}F_{jl}g^{kl}g^{ij}\right]\sqrt{g}
\]

Thus $\mathscr{R}$ is exactly the Lagrangian density for Einstein-Maxwell
theory. We get the correct ``unified'' variational principle in
a natural geometric theory without having to assume extra dimensions
(as in Kaluza-Klein theory). 

\subsubsection{Adding a Scalar Field}

If we add also a scalar potential 

\[
H=\frac{1}{2}g^{kl}\left[p_{k}-A_{k}\right]\left[p_{l}-A_{l}\right]+\phi
\]

\[
\mathcal{R}_{ij}=-g_{im}H^{k}H^{l}R_{\ klj}^{m}+\frac{1}{4}F_{ik}F_{jl}g^{kl}+\frac{1}{2}H^{k}\left\{ \partial_{j}F_{ki}+\partial_{i}F_{kj}\right\} +\partial_{i}\partial_{j}\phi
\]

In particular, the curvature of a non-relativistic particle with potential
energy $\phi$ is simply its Hessian $\partial_{i}\partial_{j}\phi$.
The harmonic oscillator has constant positive curvature. The inverted
harmonic oscillator (which has an unstable equilibrium point) has
constant negative curvature. The Ricci curvature is 

\[
\mathcal{R}=H^{k}H^{l}R_{kl}+\frac{1}{4}F_{ik}F_{jl}g^{kl}g^{ij}+H^{k}g^{ij}\partial_{i}F_{kj}+\Delta\phi
\]

Again for a a non-relativistic particle the Ricci curvature is the
Laplacian of the potential.

The generalization of the Ricci scalar density in this case 

\[
\mathfrak{R}=\left[R+\frac{1}{4}F_{ik}F_{jl}g^{kl}g^{ij}+\Delta\phi\right]e^{-\phi}\sqrt{g}
\]

Similar expression also arise as effective Lagrangian densities in
string theory and in Kaluza-Klein theories; the scalar $\phi$ is
the dilaton in that context\footnote{I thank Sumit Das for clarifying this point.}.

If we make the field redefinition

\[
\tilde{g}_{ij}=e^{2\alpha\phi}g_{ij}
\]

and choose 
\[
\alpha=-\frac{1}{n-2}
\]
 the scalar curvature density can be brought to the more conventional
form (dropping a total derivative)

\[
\mathfrak{R}=\sqrt{\tilde{g}}\tilde{R}+\frac{1}{4}F_{ik}F_{jl}\tilde{g}^{kl}\tilde{g}^{ij}\sqrt{\tilde{g}}e^{-\frac{2}{n-2}\phi}+\frac{2n-1}{n-2}\sqrt{\tilde{g}}\tilde{g}^{ij}\partial_{i}\phi\partial_{j}\phi
\]

This action describes a scalar and a photon minimally coupled to the
gravitational field, with an additional non-minimal coupling of the
scalar to the photon. The parametrization of the original Hamiltonian
is, for reference,

\[
H=\frac{1}{2}e^{-\frac{2}{n-2}\phi}\tilde{g}^{ij}[p_{i}-A_{i}][p_{j}-A_{j}]+\phi.
\]

More specific examples are given later (Section \ref{sec:Particular-Cases}).
We now turn to the explicit calculations to establish these facts.

\section{Co-ordinate transformations}

Our considerations are local, best described in old fashioned co-ordinate
notation. It is important to know how quantities transform under co-ordinate
transformations and to identify tensorial quantities, which transform
homogeneously. 

Let us begin with Hamilton's equations themselves. The configuration
space $M$ has co-ordinates $q^{i}$, which determine a canonical
co-ordinate system on $\Gamma$ with conjugate variables $p_{i}$.
We can transform to any new set of co-ordinates $\tilde{q}^{i}$ which
are smooth functions of $q^{i}$ such that the inverse transformation
is smooth as well. The momenta $\tilde{p}_{i}$ conjugate to $\tilde{q}^{i}$
are given by the transformation law of covariant vectors fields (components
of a $1-$form) 

\[
\tilde{p}_{i}=\frac{\partial q^{j}}{\partial\tilde{q}^{i}}p_{j}.
\]

It follows that $\dot{q}^{i}$ and $H^{i}=\frac{\partial H}{\partial p_{i}}$
transform as the components of a contra-variant vector field.

\[
H^{j}=\frac{\partial\tilde{q}^{j}}{\partial q^{i}}H^{i}
\]

But $\dot{p}_{i}$ and $H_{i}=\frac{\partial H}{\partial q^{i}}$
do \emph{not }transform homogeneously. Instead, 

\[
\tilde{H}_{j}=H_{b}\frac{\partial q^{b}}{\partial\tilde{q}^{j}}+H^{a}\tilde{p}_{k}\frac{\partial^{2}\tilde{q}^{k}}{\partial q^{c}\partial q^{a}}\frac{\partial q^{c}}{\partial\tilde{q}^{j}}
\]

We can see this by rewriting 

\[
dH=H_{i}dq^{i}+H^{i}dp_{i}
\]

in the new canonical co-ordinate system:

\[
dH=H_{i}\frac{\partial q^{i}}{\partial\tilde{q}^{a}}d\tilde{q}^{a}+H^{i}d\left\{ \tilde{p}_{a}\frac{\partial\tilde{q}^{a}}{\partial q^{i}}\right\} 
\]

\[
=H_{i}\frac{\partial q^{i}}{\partial\tilde{q}^{a}}d\tilde{q}^{a}+H^{i}\frac{\partial\tilde{q}^{a}}{\partial q^{i}}d\tilde{p}_{a}+H^{i}\tilde{p}_{a}\frac{\partial^{2}\tilde{q}^{a}}{\partial q^{j}\partial q^{i}}dq^{j}
\]

By collecting the coefficients of $d\tilde{p}_{j},d\tilde{q}^{j}$
we get the above transformation laws for $\tilde{H}^{j},\tilde{H}_{j}$.
It will be convenient to denote the various derivatives of the hamiltonian
by 

\[
H_{j_{1}\cdots j_{s}}^{i_{1}\cdots i_{r}}=\frac{\partial^{r+s}H}{\partial p_{i_{1}}\cdots\partial p_{i_{r}}\partial q^{j_{1}}\cdots\partial q^{j_{s}}}
\]

That is, the upper indices correspond to differentiation with respect
to $p_{i}$ and the lower indices to $q^{i}$. By extension of the
above argument, we see that $H^{i_{1}\cdots i_{r}}$ transform as
the components of a symmetric tensor under canonical co-ordinate transformations;
but that the mixed derivatives $H_{j_{1}\cdots j_{s}}^{i_{1}\cdots i_{r}}=\frac{\partial^{r+s}H}{\partial p_{i_{1}}\cdots\partial p_{i_{r}}\partial q^{j_{1}}\cdots\partial q^{j_{s}}}$
with $s>0$ transform inhomogeneously.

We are assuming that the matrix $H^{ij}$ is positive; so it has an
inverse $G_{jk}$ at every point $(q,p)$\emph{.}

\[
H^{ij}G_{jk}=\delta_{k}^{i}.
\]

This $G_{ij}$ (which could depend on $p$ as well as $q$) is our
analogue of the metric tensor; in particular it transforms covariant
tensor. \emph{But we will not use $G_{ij}$ or $H^{ij}$ to raise
or lower indices (}except when we talk of the special case of Riemannian
geometry\emph{).}
\begin{rem*}
The curvature computation makes sense if $H^{ij}$ as long as invertible,
even if not positive (as in Lorentzian geometry). It would be interesting
to generalize to the case where $H^{ij}$ are not invertible (``sub-Hamiltonian
geometry''), analogous to sub-Riemannian geometry\cite{MontgomerySubRiemGeom}.
In fact, this paper arose out of my attempts to find a formula for
curvature in sub-Riemannian geometry. 
\end{rem*}

\section{The Second Variation }

Under the variation $q^{i}\mapsto q^{i}+\epsilon\xi^{i},p_{i}\mapsto p_{i}+\epsilon\pi_{i}$
the change of the action $S=\int[p_{i}\dot{q}^{i}-H]dt$ is, to second
order, 

\[
S_{\epsilon}=S+\epsilon\int\left[\pi_{i}\dot{q}^{i}+p_{i}\dot{\xi}^{i}-H_{i}\xi^{i}-H^{i}\pi_{i}\right]dt+\epsilon^{2}\int\left[\pi_{i}\dot{\xi}^{i}-\mathcal{H}\right]dt+\mathrm{O}(\epsilon^{3})
\]

where

\[
\mathcal{H}=\frac{1}{2}\left[H_{ij}\xi^{i}\xi^{j}+2H_{i}^{j}\xi^{i}\pi_{j}+H^{ij}\pi_{i}\pi_{j}\right]
\]

Requiring that the first order variation of $S$ be zero gives us
Hamilton's equations. Given a solution of Hamilton's equations, the
second variation (``Jacobi Functional'') 

\[
\mathcal{S}=\int\left[\pi_{i}\dot{\xi}^{i}-\mathcal{H}\right]dt
\]

can be thought of as the action of a mechanical system with quadratic
(albeit time dependent) hamiltonian $\mathcal{H}$. It has an extremum
when 

\[
\dot{\xi}^{i}=H_{j}^{i}\xi^{j}+H^{ij}\pi_{j},\quad\dot{\pi}_{i}=-H_{ij}\xi^{j}-H_{i}^{j}\pi_{j}
\]

The solutions of these equations are called ``Jacobi fields''. They
describe the change of the orbit under infinitesimal perturbations
of boundary conditions. Because $H^{ij}$ is invertible, we can eliminate
$\pi_{i}$ in favor of $\dot{\xi}^{i}$ 

\[
\pi_{j}=G_{jk}\dot{\xi}^{k}-G_{jk}H_{l}^{k}\xi^{l}
\]

in the Jacobi functional to get a ``Lagrangian'' version of it:

\[
\mathcal{S}_{1}=\int\left[\frac{1}{2}G_{ij}\dot{\xi}^{i}\dot{\xi}^{j}-\dot{\xi}^{i}\xi^{j}G_{ik}H_{j}^{k}+\frac{1}{2}\xi^{i}\xi^{j}\left\{ -H_{ij}+H_{i}^{k}H_{j}^{l}G_{kl}\right\} \right]dt
\]

We will mimic standard computations of Riemannian geometry \cite{JohnLeeRiemannianManifolds}
to regroup the integrand into tensorial terms. This will lead us to
curvature.

Start with the identity 

\[
\dot{\xi}^{i}\xi^{j}=\frac{1}{2}\left[\dot{\xi}^{i}\xi^{j}-\dot{\xi}^{j}\xi^{i}\right]+\frac{1}{2}\frac{d}{dt}\left[\xi^{i}\xi^{j}\right]
\]

so that 

\[
-\dot{\xi}^{i}\xi^{j}G_{ik}H_{j}^{k}=-\frac{1}{2}\left[\dot{\xi}^{i}\xi^{j}-\dot{\xi}^{j}\xi^{i}\right]G_{ik}H_{j}^{k}-G_{ik}H_{j}^{k}\frac{1}{2}\frac{d}{dt}\left[\xi^{i}\xi^{j}\right]
\]

\[
=\frac{1}{2}\dot{\xi}^{i}\xi^{j}\left[-G_{ik}H_{j}^{k}+G_{jk}H_{i}^{k}\right]+\frac{1}{2}\xi^{i}\xi^{j}\frac{d}{dt}\left[G_{ik}H_{j}^{k}\right]+\mathrm{total\ derivative}
\]

\[
=\frac{1}{2}\dot{\xi}^{i}G_{ik}\left[-H_{j}^{k}+H^{km}G_{jl}H_{m}^{l}\right]\xi^{j}+\frac{1}{2}\xi^{i}\xi^{j}\frac{d}{dt}\left[G_{ik}H_{j}^{k}\right]+\mathrm{total\ derivative}
\]

Imitating the calculation in Riemannian geometry, we also add another
total derivative (this is the step that is not obvious):

\[
\frac{d}{dt}\left[\frac{1}{4}\dot{G}_{ij}\xi^{i}\xi^{j}\right]=\frac{1}{2}\dot{G}_{ij}\dot{\xi}^{i}\xi^{j}+\frac{1}{4}\ddot{G}_{ij}\xi^{i}\xi^{j}
\]

allowing us to write 

\[
-\dot{\xi}^{i}\xi^{j}G_{ik}H{}_{j}^{k}=\frac{1}{2}\dot{\xi}^{i}G_{ik}\left[-H_{j}^{k}+H^{km}G_{jl}H_{m}^{l}+G^{kl}\dot{G}_{lj}\right]\xi^{j}+\xi^{i}\xi^{j}\left\{ \frac{1}{2}\frac{d}{dt}\left[G_{ik}H_{j}^{k}\right]+\frac{1}{4}\ddot{G}_{ij}\right\} +\mathrm{total\ derivative}
\]

This suggests that we define an analogue of the Christoffel symbol
$\Gamma_{ij}^{k}\dot{q}^{i}$ of Riemannian geometry:

\[
\gamma_{j}^{k}=\frac{1}{2}\left[-H_{j}^{k}+H^{km}G_{jl}H_{m}^{l}+H^{kl}\dot{G}_{lj}\right]
\]

so that 

\[
-\dot{\xi}^{i}\xi^{j}G_{ik}H_{j}^{k}=\dot{\xi}^{i}G_{ik}\gamma_{j}^{k}\xi^{j}+\xi^{i}\xi^{j}\left\{ \frac{1}{2}\frac{d}{dt}\left[G_{ik}H_{j}^{k}\right]+\frac{1}{4}\ddot{G}_{ij}\right\} +\mathrm{total\ derivative}
\]

The point (which we prove below) is that although $\dot{\xi}^{i}$
does not transform as a vector , the ``covariant time derivative''

\[
\overset{\circ}{\xi}^{k}=\dot{\xi}^{k}+\gamma_{j}^{k}\xi^{j}
\]

does. \emph{We do not attempt to define a covariant derivative along
an arbitrary direction; only along the orbit of the Hamiltonian vector
field. }

We can now rewrite $\mathcal{S}_{1}$ in terms of this covariant derivative: 

\[
\mathcal{S}_{1}=\int\left[\frac{1}{2}G_{ij}\dot{\xi}^{i}\dot{\xi}^{j}+\dot{\xi}^{i}G_{ik}\gamma_{j}^{k}\xi^{j}+\xi^{i}\xi^{j}\left\{ \frac{1}{2}\frac{d}{dt}\left[G_{ik}H_{j}^{k}\right]+\frac{1}{4}\ddot{G}_{ij}\right\} +\frac{1}{2}\xi^{i}\xi^{j}\left\{ -H_{ij}+H_{i}^{k}H_{j}^{l}G_{kl}\right\} \right]dt
\]

\[
=\int\left[\frac{1}{2}G_{ij}\left\{ \dot{\xi}^{i}\dot{\xi}^{j}+2\dot{\xi}^{i}\gamma_{k}^{j}\xi^{k}\right\} +\xi^{i}\xi^{j}\left\{ \frac{1}{2}\frac{d}{dt}\left[G_{ik}H_{j}^{k}\right]+\frac{1}{4}\ddot{G}_{ij}\right\} +\frac{1}{2}\xi^{i}\xi^{j}\left\{ -H_{ij}+H_{i}^{k}H_{j}^{l}G_{kl}\right\} \right]dt
\]

\[
=\int\left[\frac{1}{2}G_{ij}\left\{ \dot{\xi}^{i}\dot{\xi}^{j}+\dot{\xi}^{i}\gamma_{k}^{j}\xi^{k}+\dot{\xi}^{j}\gamma_{k}^{i}\xi^{k}\right\} +\xi^{i}\xi^{j}\left\{ \frac{1}{2}\frac{d}{dt}\left[G_{ik}H_{j}^{k}\right]+\frac{1}{4}\ddot{G}_{ij}\right\} +\frac{1}{2}\xi^{i}\xi^{j}\left\{ -H_{ij}+H_{i}^{k}H_{j}^{l}G_{kl}\right\} \right]dt
\]

\[
=\int\left[\frac{1}{2}G_{ij}\left\{ \dot{\xi}^{i}\dot{\xi}^{j}+\dot{\xi}^{i}\gamma_{k}^{j}\xi^{k}+\gamma_{k}^{i}\xi^{k}\dot{\xi}^{j}\right\} +\xi^{i}\xi^{j}\left\{ \frac{1}{2}\frac{d}{dt}\left[G_{ik}H_{j}^{k}\right]+\frac{1}{4}\ddot{G}_{ij}\right\} +\frac{1}{2}\xi^{i}\xi^{j}\left\{ -H_{ij}+H_{i}^{k}H_{j}^{l}G_{kl}\right\} \right]dt
\]

\[
=\int\left[\frac{1}{2}G_{ij}\overset{\circ}{\xi}^{i}\overset{\circ}{\xi}^{j}-\frac{1}{2}G_{kl}\gamma_{i}^{k}\gamma_{j}^{l}\xi^{i}\xi^{j}+\xi^{i}\xi^{j}\left\{ \frac{1}{2}\frac{d}{dt}\left[G_{ik}H_{j}^{k}\right]+\frac{1}{4}\ddot{G}_{ij}\right\} +\frac{1}{2}\xi^{i}\xi^{j}\left\{ -H_{ij}+H_{i}^{k}H_{j}^{l}G_{kl}\right\} \right]dt
\]

Thus

\[
\mathcal{S}=\int\left[\frac{1}{2}G_{ij}\overset{\circ}{\xi}^{i}\overset{\circ}{\xi}^{j}-\xi^{i}\xi^{j}\frac{1}{2}\left\{ G_{kl}\gamma_{i}^{k}\gamma_{j}^{l}-\frac{d}{dt}\left[G_{ik}H_{j}^{k}\right]-\frac{1}{2}\ddot{G}_{ij}+H_{ij}-H_{i}^{k}H_{j}^{l}G_{kl}\right\} \right]dt
\]

The symmetric part of the quantity in the curly brackets is an analogue
of curvature.  

\[
\mathcal{S}_{1}=\int\left[\frac{1}{2}G_{ij}\overset{\circ}{\xi}^{i}\overset{\circ}{\xi}^{j}-\frac{1}{2}\xi^{i}\xi^{j}\mathcal{R}_{ij}\right]dt
\]

Rewriting time derivatives as Poisson Brackets, we can express it
in terms of the first four derivatives of the hamiltonian:

\begin{equation}
\mathcal{R}_{ij}=G_{kl}\gamma_{i}^{k}\gamma_{j}^{l}-\frac{1}{2}\left\{ H,G_{ik}H_{j}^{k}+G_{jk}H_{i}^{k}\right\} -\frac{1}{2}\left\{ H,\left\{ H,G_{ij}\right\} \right\} +H_{ij}-H_{i}^{k}H_{j}^{l}G_{kl}\label{eq:Curvature}
\end{equation}

The trace 
\begin{quote}
\[
\mathcal{R}=H^{ij}\mathcal{R}_{ij}
\]

plays the role of the Ricci tensor. There is no notion of Ricci scalar
in general Hamiltonian mechanics. But, it makes sense to integrate
this w.r.t. the Boltzmann measure

\[
\mathfrak{R}(H)=\int\mathcal{R}\ d\mu_{H}
\]

to give a functional of the hamiltonian. We will see that this reduces
to the integral of the Ricci scalar over a Riemannian manifold.
\end{quote}

\subsection{The Transformation of $\gamma_{j}^{i}$ and $\mathcal{R}_{ij}$}

Recall that $\xi^{i}$ transforms as the components of a vector field:

\[
\tilde{\xi}^{i}=\frac{\partial\tilde{q}^{i}}{\partial q^{j}}\ \xi^{j}
\]

but not its time derivative:

\[
\frac{d}{dt}\tilde{\xi}^{i}=\frac{d}{dt}\left[\frac{\partial\tilde{q}^{i}}{\partial q^{j}}\xi^{j}\right]
\]

\[
\dot{\tilde{\xi}}^{i}=\frac{\partial\tilde{q}^{i}}{\partial q^{j}}\dot{\xi}^{j}+\frac{\partial^{2}\tilde{q}^{i}}{\partial q^{k}\partial q^{j}}\dot{q}^{k}\xi^{j}
\]

The inhomogeneous term in the covariant time derivative 

\[
\overset{\circ}{\xi}^{i}=\dot{\xi}^{i}+\gamma_{j}^{k}\xi^{j}
\]

is cancelled if the symbols $\gamma_{j}^{i}$ transform as 

\[
\tilde{\gamma}_{j}^{i}=\frac{\partial\tilde{q}^{i}}{\partial q^{k}}\gamma_{l}^{k}\frac{\partial q^{l}}{\partial\tilde{q}^{j}}-\frac{\partial q^{l}}{\partial\tilde{q}^{j}}\frac{\partial^{2}\tilde{q}^{i}}{\partial q^{k}\partial q^{l}}\dot{q}^{k}
\]

For,

\[
\overset{\circ}{\tilde{\xi}}^{i}=\dot{\tilde{\xi}}^{i}+\tilde{\gamma}_{j}^{i}\tilde{\xi}^{j}
\]

\[
=\frac{\partial\tilde{q}^{i}}{\partial q^{j}}\dot{\xi}^{j}+\frac{\partial^{2}\tilde{q}^{i}}{\partial q^{k}\partial q^{j}}\dot{q}^{k}\xi^{j}+\left\{ \frac{\partial\tilde{q}^{i}}{\partial q^{k}}\gamma_{l}^{k}\frac{\partial q^{l}}{\partial\tilde{q}^{j}}-\frac{\partial q^{l}}{\partial\tilde{q}^{j}}\frac{\partial^{2}\tilde{q}^{i}}{\partial q^{k}\partial q^{l}}\dot{q}^{k}\right\} \frac{\partial\tilde{q}^{j}}{\partial q^{m}}\ \xi^{m}
\]

\[
=\frac{\partial\tilde{q}^{i}}{\partial q^{j}}\left[\dot{\xi}^{j}+\gamma_{l}^{j}\xi^{l}\right]
\]

Using Hamilton's equations, we can write the required transformation
law as 

\[
\tilde{\gamma}_{j}^{i}=\frac{\partial\tilde{q}^{i}}{\partial q^{k}}\gamma_{l}^{k}\frac{\partial q^{l}}{\partial\tilde{q}^{j}}-\frac{\partial q^{l}}{\partial\tilde{q}^{j}}\frac{\partial^{2}\tilde{q}^{i}}{\partial q^{k}\partial q^{l}}H^{k}
\]

For later reference we rewrite this by a relabelling of indices as 

\[
\tilde{\gamma}_{k}^{l}=\frac{\partial q^{c}}{\partial\tilde{q}^{k}}\frac{\partial\tilde{q}^{l}}{\partial q^{a}}\gamma_{c}^{a}-\frac{\partial q^{c}}{\partial\tilde{q}^{k}}\frac{\partial^{2}\tilde{q}^{l}}{\partial q^{c}\partial q^{a}}H^{a}.
\]

\begin{prop*}
$\gamma_{k}^{l}$ transforms as above. So $\overset{\circ}{\xi}^{i}$
and $\mathcal{R}_{ij}$ transform as tensors.
\end{prop*}
\textbf{Proof }We need the transformations of $H_{k}^{l},P_{k}^{l}\equiv H^{lj}G_{ki}H_{j}^{i},G_{kj}\dot{H}^{ik}$.

\subsubsection*{The Transformation of $H_{k}^{l}$}

Recall that 

\[
\tilde{H}_{j}=H_{b}\frac{\partial q^{b}}{\partial\tilde{q}^{j}}+H^{a}\tilde{p}_{k}\frac{\partial^{2}\tilde{q}^{k}}{\partial q^{c}\partial q^{a}}\frac{\partial q^{c}}{\partial\tilde{q}^{j}}
\]

By differentiating w.r.t. to $\tilde{p}_{i}$ we get the transformation
of $\tilde{H}_{j}^{i}$:

\[
\tilde{H}_{j}^{i}=\frac{\partial q^{b}}{\partial\tilde{q}^{j}}\frac{\partial\tilde{q}^{i}}{\partial q^{a}}H_{b}^{a}+\frac{\partial q^{c}}{\partial\tilde{q}^{j}}\frac{\partial^{2}\tilde{q}^{i}}{\partial q^{c}\partial q^{a}}H^{a}+\frac{\partial q^{c}}{\partial\tilde{q}^{j}}\ \frac{\partial\tilde{q}^{i}}{\partial q^{a}}\ \frac{\partial^{2}\tilde{q}^{k}}{\partial q^{c}\partial q^{b}}H^{ab}\tilde{p}_{k}
\]

Relabelling indices (for later use)

\[
\tilde{H}_{k}^{l}=\frac{\partial q^{c}}{\partial\tilde{q}^{k}}\frac{\partial\tilde{q}^{l}}{\partial q^{a}}H_{c}^{a}+\frac{\partial q^{c}}{\partial\tilde{q}^{k}}\frac{\partial^{2}\tilde{q}^{l}}{\partial q^{c}\partial q^{a}}H^{a}+\frac{\partial q^{b}}{\partial\tilde{q}^{k}}\ \frac{\partial\tilde{q}^{l}}{\partial q^{d}}\ \frac{\partial^{2}\tilde{q}^{m}}{\partial q^{b}\partial q^{c}}H^{dc}\tilde{p}_{m}
\]

\subsubsection*{The Transformation of $P_{k}^{l}$}

\[
\tilde{P}_{k}^{l}=\tilde{H}^{lj}\tilde{G}_{ki}\left\{ \frac{\partial q^{b}}{\partial\tilde{q}^{j}}\frac{\partial\tilde{q}^{i}}{\partial q^{a}}H_{b}^{a}+\frac{\partial q^{c}}{\partial\tilde{q}^{j}}\frac{\partial^{2}\tilde{q}^{i}}{\partial q^{c}\partial q^{a}}H^{a}+\frac{\partial q^{c}}{\partial\tilde{q}^{j}}\ \frac{\partial\tilde{q}^{i}}{\partial q^{a}}\ \frac{\partial^{2}\tilde{q}^{m}}{\partial q^{c}\partial q^{b}}H^{ab}\tilde{p}_{m}\right\} 
\]

Note that 

\[
\tilde{H}^{lj}\tilde{G}_{ki}\frac{\partial q^{c}}{\partial\tilde{q}^{j}}\ \frac{\partial\tilde{q}^{i}}{\partial q^{a}}=H^{dc}G_{na}\frac{\partial\tilde{q}^{l}}{\partial q^{d}}\frac{\partial q^{n}}{\partial\tilde{q}^{k}}
\]

so that 

\[
\tilde{P}_{k}^{l}=H^{mb}G_{na}H_{b}^{a}\frac{\partial\tilde{q}^{l}}{\partial q^{m}}\frac{\partial q^{n}}{\partial\tilde{q}^{k}}+\frac{\partial\tilde{q}^{l}}{\partial q^{m}}H^{mc}\tilde{G}_{ki}\frac{\partial^{2}\tilde{q}^{i}}{\partial q^{c}\partial q^{a}}H^{a}+H^{dc}G_{na}\frac{\partial\tilde{q}^{l}}{\partial q^{d}}\frac{\partial q^{n}}{\partial\tilde{q}^{k}}\frac{\partial^{2}\tilde{q}^{m}}{\partial q^{c}\partial q^{b}}H^{ab}\tilde{p}_{m}
\]

\[
=P_{n}^{m}\frac{\partial\tilde{q}^{l}}{\partial q^{m}}\frac{\partial q^{n}}{\partial\tilde{q}^{k}}+\frac{\partial\tilde{q}^{l}}{\partial q^{m}}H^{mc}\tilde{G}_{ki}\frac{\partial^{2}\tilde{q}^{i}}{\partial q^{c}\partial q^{a}}H^{a}+H^{dc}\frac{\partial\tilde{q}^{l}}{\partial q^{d}}\frac{\partial q^{b}}{\partial\tilde{q}^{k}}\frac{\partial^{2}\tilde{q}^{m}}{\partial q^{c}\partial q^{b}}\tilde{p}_{m}
\]

\subsubsection*{The Transformation of $G_{ki}\dot{H}^{il}$}

\[
\tilde{G}_{ki}\dot{\tilde{H}}^{il}=\tilde{G}_{ki}\frac{d}{dt}\left[\frac{\partial\tilde{q}^{i}}{\partial q^{a}}\frac{\partial\tilde{q}^{l}}{\partial q^{c}}H^{ac}\right]
\]

\[
=\tilde{G}_{ki}\frac{\partial\tilde{q}^{i}}{\partial q^{a}}\frac{\partial\tilde{q}^{l}}{\partial q^{c}}\dot{H}^{ac}+\tilde{G}_{ki}\frac{\partial\tilde{q}^{i}}{\partial q^{a}}\frac{\partial^{2}\tilde{q}^{l}}{\partial q^{c}\partial q^{b}}\dot{q}^{b}H^{ac}+\tilde{G}_{ki}\frac{\partial^{2}\tilde{q}^{i}}{\partial q^{a}\partial q^{b}}\frac{\partial\tilde{q}^{l}}{\partial q^{c}}\dot{q}^{b}H^{ac}
\]

\[
=\tilde{G}_{ki}\frac{\partial\tilde{q}^{i}}{\partial q^{a}}\frac{\partial\tilde{q}^{l}}{\partial q^{c}}\dot{H}^{ac}+\frac{\partial^{2}\tilde{q}^{l}}{\partial q^{c}\partial q^{b}}\dot{q}^{b}\left\{ H^{ac}\tilde{G}_{ki}\frac{\partial\tilde{q}^{i}}{\partial q^{a}}\right\} +\tilde{G}_{ki}\frac{\partial^{2}\tilde{q}^{i}}{\partial q^{a}\partial q^{b}}\frac{\partial\tilde{q}^{l}}{\partial q^{c}}\dot{q}^{b}H^{ac}
\]

\[
=\tilde{G}_{ki}\frac{\partial\tilde{q}^{i}}{\partial q^{a}}\frac{\partial\tilde{q}^{l}}{\partial q^{c}}H^{ac}+\frac{\partial^{2}\tilde{q}^{l}}{\partial q^{c}\partial q^{b}}\dot{q}^{b}\left\{ H^{ac}G_{ad}\frac{\partial q^{d}}{\partial\tilde{q}^{k}}\right\} +\tilde{G}_{ki}\frac{\partial^{2}\tilde{q}^{i}}{\partial q^{a}\partial q^{b}}\frac{\partial\tilde{q}^{l}}{\partial q^{c}}\dot{q}^{b}H^{ac}
\]

\[
=\tilde{G}_{ki}\frac{\partial\tilde{q}^{i}}{\partial q^{a}}\frac{\partial\tilde{q}^{l}}{\partial q^{c}}\dot{H}^{ac}+\frac{\partial^{2}\tilde{q}^{l}}{\partial q^{c}\partial q^{b}}\dot{q}^{b}\frac{\partial q^{c}}{\partial\tilde{q}^{k}}+\tilde{G}_{ki}\frac{\partial^{2}\tilde{q}^{i}}{\partial q^{a}\partial q^{b}}\frac{\partial\tilde{q}^{l}}{\partial q^{c}}\dot{q}^{b}H^{ac}
\]

\[
=\tilde{G}_{ki}\frac{\partial\tilde{q}^{i}}{\partial q^{a}}\frac{\partial\tilde{q}^{l}}{\partial q^{c}}\dot{H}^{ac}+\frac{\partial^{2}\tilde{q}^{l}}{\partial q^{c}\partial q^{a}}\frac{\partial q^{c}}{\partial\tilde{q}^{k}}\dot{q}^{a}+\tilde{G}_{ki}\frac{\partial^{2}\tilde{q}^{i}}{\partial q^{a}\partial q^{b}}\frac{\partial\tilde{q}^{l}}{\partial q^{c}}\dot{q}^{b}H^{ac}
\]

\[
=\tilde{G}_{ki}\frac{\partial\tilde{q}^{i}}{\partial q^{a}}\frac{\partial\tilde{q}^{l}}{\partial q^{c}}\dot{H}^{ac}+\frac{\partial^{2}\tilde{q}^{l}}{\partial q^{c}\partial q^{a}}\frac{\partial q^{c}}{\partial\tilde{q}^{k}}H^{a}+\tilde{G}_{ki}\frac{\partial^{2}\tilde{q}^{i}}{\partial q^{a}\partial q^{b}}\frac{\partial\tilde{q}^{l}}{\partial q^{c}}H^{a}H^{ac}
\]

\[
=\frac{\partial q^{b}}{\partial\tilde{q}^{k}}\frac{\partial\tilde{q}^{l}}{\partial q^{c}}G_{ba}\dot{H}^{ac}+\frac{\partial^{2}\tilde{q}^{l}}{\partial q^{c}\partial q^{a}}\frac{\partial q^{c}}{\partial\tilde{q}^{k}}H^{a}+\tilde{G}_{ki}\frac{\partial^{2}\tilde{q}^{i}}{\partial q^{a}\partial q^{b}}\frac{\partial\tilde{q}^{l}}{\partial q^{c}}H^{a}H^{ac}
\]

where, we use Hamilton's equation $\dot{q}^{a}=H^{a}$.

Relabeling $c\to a,a\to b,b\to c$ in the first term and $c\to m,a\to c,b\to a,$
in the last term (for later use),

\[
\tilde{G}_{ki}\dot{\tilde{H}}^{il}=\frac{\partial q^{c}}{\partial\tilde{q}^{k}}\frac{\partial\tilde{q}^{l}}{\partial q^{a}}G_{cb}\dot{H}^{ba}+\frac{\partial^{2}\tilde{q}^{l}}{\partial q^{c}\partial q^{a}}\frac{\partial q^{c}}{\partial\tilde{q}^{k}}H^{a}+\tilde{G}_{ki}\frac{\partial^{2}\tilde{q}^{i}}{\partial q^{c}\partial q^{a}}\frac{\partial\tilde{q}^{l}}{\partial q^{m}}H^{a}H^{cm}
\]

So consider the linear combination 

\[
A\tilde{H}_{k}^{l}+B\tilde{P}_{k}^{l}+C\tilde{G}_{ki}\dot{\tilde{G}}^{il}
\]

\[
=A\left\{ \frac{\partial q^{c}}{\partial\tilde{q}^{k}}\frac{\partial\tilde{q}^{l}}{\partial q^{a}}H_{c}^{a}+\frac{\partial q^{c}}{\partial\tilde{q}^{k}}\frac{\partial^{2}\tilde{q}^{l}}{\partial q^{c}\partial q^{a}}H^{a}+\frac{\partial q^{b}}{\partial\tilde{q}^{k}}\ \frac{\partial\tilde{q}^{l}}{\partial q^{d}}\ \frac{\partial^{2}\tilde{q}^{m}}{\partial q^{b}\partial q^{c}}H^{dc}\tilde{p}_{m}\right\} 
\]

\[
+B\left\{ P_{c}^{a}\frac{\partial\tilde{q}^{l}}{\partial q^{a}}\frac{\partial q^{c}}{\partial\tilde{q}^{k}}+\frac{\partial\tilde{q}^{l}}{\partial q^{m}}H^{mc}\tilde{G}_{ki}\frac{\partial^{2}\tilde{q}^{i}}{\partial q^{c}\partial q^{a}}H^{a}+H^{dc}\frac{\partial\tilde{q}^{l}}{\partial q^{d}}\frac{\partial q^{b}}{\partial\tilde{q}^{k}}\frac{\partial^{2}\tilde{q}^{m}}{\partial q^{c}\partial q^{b}}\tilde{p}_{m}\right\} 
\]

\[
+C\left\{ \frac{\partial q^{c}}{\partial\tilde{q}^{k}}\frac{\partial\tilde{q}^{l}}{\partial q^{a}}G_{cb}\dot{H}^{ba}+\frac{\partial^{2}\tilde{q}^{l}}{\partial q^{c}\partial q^{a}}\frac{\partial q^{c}}{\partial\tilde{q}^{k}}H^{a}+\tilde{G}_{ki}\frac{\partial^{2}\tilde{q}^{i}}{\partial q^{c}\partial q^{a}}\frac{\partial\tilde{q}^{l}}{\partial q^{m}}H^{a}H^{cm}\right\} 
\]

If we choose 

\[
A+C=-1,\quad B+C=0,\quad A+B=0\implies A=-\frac{1}{2}=C,\quad B=\frac{1}{2}
\]

we get the transformation law 

\[
\frac{1}{2}\left[-\tilde{H}_{k}^{l}+\tilde{P}_{k}^{l}-\tilde{G}_{ki}\dot{\tilde{G}}^{il}\right]=\frac{\partial q^{c}}{\partial\tilde{q}^{k}}\frac{\partial\tilde{q}^{l}}{\partial q^{a}}\frac{1}{2}\left[-H_{c}^{a}+P_{c}^{a}-G_{cb}\dot{H}^{ba}\right]-\frac{\partial q^{c}}{\partial\tilde{q}^{k}}\frac{\partial^{2}\tilde{q}^{l}}{\partial q^{c}\partial q^{a}}H^{a}
\]

So we can choose 

\[
\gamma_{c}^{a}=\frac{1}{2}\left[-H_{c}^{a}+P_{c}^{a}-G_{cb}\dot{H}^{ba}\right]=\frac{1}{2}\left[-H_{c}^{a}+P_{c}^{a}+\dot{G}_{cb}H^{ba}\right]
\]

(where we used $\dot{G}_{cb}H^{ba}+G_{cb}\dot{H}^{ba}=0$) to get
the transformation law 

\[
\tilde{\gamma}_{k}^{l}=\frac{\partial q^{c}}{\partial\tilde{q}^{k}}\frac{\partial\tilde{q}^{l}}{\partial q^{a}}\gamma_{c}^{a}-\frac{\partial q^{c}}{\partial\tilde{q}^{k}}\frac{\partial^{2}\tilde{q}^{l}}{\partial q^{c}\partial q^{a}}H^{a}
\]

This is what we wanted.

\section{Comparison With Riemannian Geometry}

We must show that the above formula (\ref{eq:Curvature}) reduces
to the usual one for curvature in Riemannian geometry. In this section,
unlike before, we raise and lower indices using the metric tensor.

If $H=\frac{1}{2}g^{ij}p_{i}p_{j}$ Hamilton's equations reduce to
the geodesic equation\cite{JohnLeeRiemannianManifolds} . Also, 

\[
H^{ij}=g^{ij}
\]

\[
H_{i}^{j}=\partial_{i}g^{jk}p_{k}
\]

Recalling the formula for the Christoffel symbols, 

\[
\Gamma_{jk}^{i}=\frac{1}{2}g^{im}\left[\partial_{k}g_{jm}+\partial_{j}g_{km}-\partial_{m}g_{jk}\right]
\]

we can rewrite them in terms of the contra-variant metric tensor

\[
\Gamma_{jk}^{i}\dot{q}^{k}=\frac{1}{2}g^{im}\left[\partial_{k}g_{jm}+\partial_{j}g_{km}-\partial_{m}g_{jk}\right]\dot{q}^{k}
\]

\[
=\frac{1}{2}g^{im}\left[\dot{g}_{jm}+\left\{ -g_{ka}g_{mb}\partial_{j}g^{ab}+g_{ja}g_{kb}\partial_{m}g^{ab}\right\} \dot{q}^{k}\right]
\]

\[
=\frac{1}{2}\left[g^{ik}\dot{g}_{jk}+\left\{ -p_{a}\partial_{j}g^{ai}+g_{ja}p_{b}g^{im}\partial_{m}g^{ab}\right\} \right]
\]

This is a particular case of the general formula 

\[
\gamma_{j}^{i}=\frac{1}{2}\left[H^{ik}\dot{G}_{kj}-H_{j}^{i}+G_{jl}H_{m}^{l}H^{im}\right]
\]

so that $\gamma_{j}^{i}$ reduces to $\Gamma_{jk}^{i}\dot{q}^{k}$
in Riemannian geometry. 

The Riemann tensor is

\[
R_{ijk}^{l}=\partial_{j}\Gamma_{ik}^{l}-\partial_{k}\Gamma_{ij}^{l}+\Gamma_{jm}^{l}\Gamma_{ik}^{m}-\Gamma_{km}^{l}\Gamma_{ij}^{m}
\]

\[
R_{lijk}=g_{ln}R_{\ ijk}^{n}
\]

To compare the curvatures, it is convenient to choose Riemann normal
co-ordinates in which the first derivative of the metric is zero at
the chosen point.( $\approx$ denotes equality in Riemannian normal
co-ordinates up to higher order terms.)

\[
g_{ij}\approx\delta_{ij}
\]

\[
R_{iklm}\approx\frac{1}{2}\left(\partial_{k}\partial_{l}g_{im}+\partial_{i}\partial_{m}g_{kl}-\partial_{k}\partial_{m}g_{il}-\partial_{i}\partial_{l}g_{km}\right)
\]

\[
R_{iklj}\approx\frac{1}{2}\left(\partial_{k}\partial_{l}g_{ij}+\partial_{i}\partial_{j}g_{kl}-\partial_{k}\partial_{j}g_{il}-\partial_{i}\partial_{l}g_{kj}\right)
\]

On the other hand,

\[
\mathcal{R}_{ij}\approx-\frac{1}{2}\left(\dot{H}_{j}^{i}+\dot{H}_{i}^{j}\right)-\frac{1}{2}\ddot{G}_{ij}+H_{ij}
\]

In Riemannian geometry,

\[
\dot{G}_{ij}\equiv\left\{ H,G_{ij}\right\} =H^{k}\partial_{k}g_{ij}
\]

\[
\ddot{G}_{ij}=H^{m}\partial_{m}\left[H^{n}\partial_{n}g_{ij}\right]-H_{m}\frac{\partial}{\partial p_{m}}\left[H^{n}\partial_{n}g_{ij}\right]=H^{m}H^{n}\partial_{m}\partial_{n}g_{ij}+H^{m}H_{m}^{n}\partial_{n}g_{ij}-H_{m}H^{mn}\partial_{n}g_{ij}
\]

\[
\dot{H}_{j}^{i}\equiv\left\{ H,\partial_{i}g^{jk}p_{k}\right\} =H^{m}\partial_{m}\partial_{i}g^{jk}p_{k}-H_{k}\partial_{i}g^{jk}
\]

so that in normal co-ordinates

\[
\ddot{G}_{ij}\approx p_{k}p_{l}\partial_{k}\partial_{l}g_{ij}
\]

\[
\dot{H}_{j}^{i}\approx p_{k}p_{m}\partial_{m}\partial_{i}g^{jk}\approx-p_{k}p_{l}\partial_{l}\partial_{i}g_{jk}
\]

\[
H_{ij}=\frac{1}{2}\partial_{i}\partial_{j}g^{kl}p_{k}p_{l}\approx-\frac{1}{2}p_{k}p_{l}\partial_{i}\partial_{j}g_{kl}
\]

\[
\mathcal{R}_{ij}\approx\frac{1}{2}p_{k}p_{l}\left(\partial_{l}\partial_{i}g_{jk}+\partial_{l}\partial_{j}g_{ik}\right)-\frac{1}{2}p_{k}p_{l}\partial_{k}\partial_{l}g_{ij}-\frac{1}{2}p_{k}p_{l}\partial_{i}\partial_{j}g_{kl}
\]

Thus 

\[
\mathcal{R}_{ij}\approx-p_{k}p_{l}R_{iklj}
\]

Since $\mathcal{R}_{ij}$ and $R_{ijk}^{l}$ are tensors we get the
equality in general co-ordinates

\[
\mathcal{R}_{ij}=-g_{im}H^{k}H^{l}R_{\ klj}^{m},\quad H^{m}=g^{mn}p_{n}
\]

Moreover

\[
H^{ij}\mathcal{R}_{ij}=-H^{k}H^{l}R_{\ klm}^{m}=H^{k}H^{l}R_{\ kml}^{m}
\]

so that 

\[
\mathcal{R}\equiv H^{ij}\mathcal{R}_{ij}=H^{k}H^{l}R_{kl}.
\]

Note the Gaussian integrals 

\[
\int e^{-\frac{1}{2}g^{ij}p_{i}p_{j}}\frac{d^{n}p}{\left[2\pi\right]^{\frac{n}{2}}}=\sqrt{g},\quad\int e^{-\frac{1}{2}g^{ij}p_{i}p_{j}}p_{k}p_{l}\frac{d^{n}p}{\left[2\pi\right]^{\frac{n}{2}}}=\sqrt{g}g_{kl}
\]

Thus we get the Einstein-Hilbert Lagrangian for GR as the Boltzmann
average of the Ricci tensor:

\[
\mathscr{R}\equiv\int\mathcal{R}\ e^{-H}\frac{d^{n}p}{\left[2\pi\right]^{\frac{n}{2}}}=R\sqrt{g},\quad R=R_{kl}g^{kl}.
\]

Thus there might be some merit in considering $\mathscr{R}(H)$ as
a variational principle that determines the hamiltonian itself in
the general case. 

\section{Adding a Magnetic Field and a Scalar Potential}

The typical hamiltonian of a point particle in physics is a polynomial
of order two in the momenta; it describes its interaction with a gravitational
electromagnetic and scalar field 

\[
H=\frac{1}{2}g^{kl}\left[p_{k}-A_{k}\right]\left[p_{l}-A_{l}\right]+\phi
\]

We can compute,

\[
H^{k}=g^{kl}\left[p_{l}-A_{l}\right]
\]

\[
H_{j}^{k}=\partial_{j}g^{kl}\left[p_{l}-A_{l}\right]-g^{kl}\partial_{j}A_{l}
\]

\[
H_{m}^{l}=\partial_{m}g^{ln}\left[p_{n}-A_{n}\right]-g^{ln}\partial_{m}A_{n}
\]

\[
H^{kl}=g^{kl},\quad G_{kl}=g_{kl}
\]

\[
H^{km}G_{jl}H_{m}^{l}=g^{km}g_{jl}\left\{ \partial_{m}g^{ln}\left[p_{n}-A_{n}\right]-g^{ln}\partial_{m}A_{n}\right\} 
\]

\[
=-g^{km}\left[\partial_{m}g_{jl}\right]g^{ln}\left[p_{n}-A_{n}\right]-g^{km}\partial_{m}A_{j}
\]

\[
H^{kl}\dot{G}_{ij}=g^{kl}H^{n}\partial_{n}g_{kl}
\]
To proceed further we pass to the Riemann normal co-ordinates normal
co-ordinates; also choose $A_{i}=0$ at the origin by a choice of
gauge (but of course not the derivative $\partial_{i}A_{j}$).

\[
\gamma_{j}^{k}\approx\frac{1}{2}\left[\partial_{j}A_{k}-\partial_{k}A_{j}\right]=\frac{1}{2}F_{jk}
\]

and

\[
\mathcal{R}_{ij}\approx\gamma_{i}^{k}\gamma_{j}^{k}-\frac{1}{2}\left\{ \dot{H}_{j}^{i}+\dot{H}_{i}^{j}\right\} -\frac{1}{2}\ddot{G}_{ij}+H_{ij}-H_{i}^{k}H_{j}^{l}G_{kl}
\]

\[
\dot{H}_{j}^{i}\approx-p_{k}p_{l}\partial_{l}\partial_{i}g_{jk}-p_{k}\partial_{k}\partial_{j}A_{i}
\]

\[
\ddot{G}_{ij}\approx p_{k}p_{l}\partial_{k}\partial_{l}g_{ij}
\]

\[
H_{ij}\approx-\frac{1}{2}p_{k}p_{l}\partial_{i}\partial_{j}g_{kl}+g^{kl}\partial_{i}A_{k}\partial_{j}A_{l}-p_{k}\partial_{i}\partial_{j}A_{k}+\partial_{i}\partial_{j}\phi
\]

\[
H_{i}^{k}H_{j}^{l}G_{kl}\approx\partial_{i}A_{k}\partial_{j}A_{k}
\]

\[
-\frac{1}{2}\left\{ \dot{H}_{j}^{i}+\dot{H}_{i}^{j}\right\} +H_{ij}-H_{i}^{k}H_{j}^{l}G_{kl}\approx\frac{1}{2}p_{k}p_{l}\left[\partial_{l}\partial_{i}g_{jk}+\partial_{l}\partial_{j}g_{ik}-\partial_{i}\partial_{j}g_{kl}\right]+
\]

\[
p_{k}\left\{ \frac{1}{2}\partial_{k}\partial_{j}A_{i}+\frac{1}{2}\partial_{k}\partial_{i}A_{j}-\partial_{i}\partial_{j}A_{k}\right\} +g^{kl}\partial_{i}A_{k}\partial_{j}A_{l}-\partial_{i}A_{k}\partial_{j}A_{k}+\partial_{i}\partial_{j}\phi
\]

\[
=\frac{1}{2}p_{k}p_{l}\left[\partial_{l}\partial_{i}g_{jk}+\partial_{l}\partial_{j}g_{ik}-\partial_{i}\partial_{j}g_{kl}\right]+\frac{1}{2}p_{k}\left\{ \partial_{j}F_{ki}+\partial_{i}F_{kj}\right\} +\partial_{i}\partial_{j}\phi
\]

\[
\mathcal{R}_{ij}\approx\frac{1}{4}F_{ik}F_{jk}+\frac{1}{2}p_{k}p_{l}\left[\partial_{l}\partial_{i}g_{jk}+\partial_{l}\partial_{j}g_{ik}-\partial_{i}\partial_{j}g_{kl}\right]+\frac{1}{2}p_{k}\left\{ \partial_{j}F_{ki}+\partial_{i}F_{kj}\right\} +\partial_{i}\partial_{j}\phi
\]

In a general co-ordinate system, this is the tensorial equality

\[
\mathcal{R}_{ij}=-g_{im}H^{k}H^{l}R_{\ klj}^{m}+\frac{1}{4}F_{ik}F_{jl}g^{kl}+\frac{1}{2}H^{k}\left\{ \partial_{j}F_{ki}+\partial_{i}F_{kj}\right\} +\partial_{i}\partial_{j}\phi
\]

Taking a trace

\[
\mathcal{R}=H^{k}H^{l}R_{kl}+\frac{1}{4}F_{ik}F_{jl}g^{kl}g^{ij}+H^{k}g^{ij}\partial_{i}F_{kj}+\Delta\phi
\]

\subsection{An Action Principle for Fields}

The integral over momentum with the Boltzmann weight now has an extra
factor of $e^{-\phi}$:

\[
\int e^{-\left\{ \frac{1}{2}g^{ij}\left[p_{i}-A_{i}\right]\left[p_{j}-A_{j}\right]+\phi\right\} }\frac{d^{n}p}{\left[2\pi\right]^{\frac{n}{2}}}=e^{-\phi}\sqrt{g},\quad
\]

\[
\int e^{-\left\{ \frac{1}{2}g^{ij}\left[p_{i}-A_{i}\right]\left[p_{j}-A_{j}\right]+\phi\right\} }\left[p_{k}-A_{k}\right]\left[p_{l}-A_{l}\right]\frac{d^{n}p}{\left[2\pi\right]^{\frac{n}{2}}}=e^{-\phi}\sqrt{g}g_{kl}
\]

Thus 

\[
\mathscr{R}\equiv\int\mathcal{R}\ e^{-H}\frac{d^{n}p}{\left[2\pi\right]^{\frac{n}{2}}}=\left[R+\frac{1}{4}F_{ik}F_{jl}g^{kl}g^{ij}+\Delta\phi\right]e^{-\phi}\sqrt{g}
\]

Similar actions also arise in string theory and in Kaluza-Klein theories;
the scalar $\phi$ is the dilaton in that context.

If we make the field redefinition

\[
\tilde{g}_{ij}=e^{2\alpha\phi}g_{ij}
\]

and choose 
\[
\alpha=-\frac{1}{n-2}
\]
 the action of the fields can be brought to the more conventional
form (dropping a total derivative)

\[
\mathscr{R}=\sqrt{\tilde{g}}\tilde{R}+\frac{1}{4}F_{ik}F_{jl}\tilde{g}^{kl}\tilde{g}^{ij}\sqrt{\tilde{g}}e^{-\frac{2}{n-2}\phi}+\frac{2n-1}{n-2}\sqrt{\tilde{g}}\tilde{g}^{ij}\partial_{i}\phi\partial_{j}\phi
\]

The parametrization of the original Hamiltonian is, for reference,

\[
H=\frac{1}{2}e^{-\frac{2}{n-2}\phi}\tilde{g}^{ij}[p_{i}-A_{i}][p_{j}-A_{j}]+\phi.
\]

This action describes a scalar and a photon minimally coupled to the
gravitational field, with an additional non-minimal coupling of the
scalar to the photon.

 \textbf{}

\section{Particular Cases\label{sec:Particular-Cases}}

\subsection{The Free Particle}

The simplest, but rather trivial, case is the hamiltonian of a particle
moving on the real line 

\[
H=\frac{1}{2}p^{2}
\]

Of course, the curvature is zero. The trajectories are straight lines.
It is straightforward to get

\[
s_{T}(Q,Q')=\frac{(Q-Q')^{2}}{2T},\quad\sigma_{E}(Q,Q')=\sqrt{2E}|Q-Q'|
\]

They satisfy the HJ equations

\[
\frac{1}{2}\left[\partial_{Q}s_{T}\right]^{2}+\frac{\partial s_{T}}{\partial T}=0,\quad\frac{1}{2}\left[\partial_{Q}\sigma_{E}\right]^{2}=E
\]

\subsection{The Harmonic Oscillator}

The simplest non-Euclidean geometry is the sphere; it has constant
positive curvature. The mechanical analogue is the simple harmonic
oscillator

\[
H=\frac{1}{2}\left[p^{2}+\omega^{2}q^{2}\right]
\]

The Lagrangian sub-manifold (configuration space) $M$ is one-dimensional,
just the real line $\mathbb{R}$. Even the real line is curved in
our sense! It is constant and positive:

\[
\mathcal{R}_{11}=\omega^{2}
\]

In this is case, this is also the Ricci form. The phase space has
finite volume in the Boltzmann measure; the induced volume element
on the real line is the Gaussian.

\[
dq\int e^{-H}\frac{dp}{\sqrt{2\pi}}=e^{-\frac{1}{2}\omega^{2}q^{2}}dq.
\]

Given $E>0$, the set of allowed positions 

\[
M_{E}=\left\{ q\mid H(q,p)=E\mathrm{\ for\ some\ }p\right\} =\left[-\frac{\sqrt{2E}}{\omega},\frac{\sqrt{2E}}{\omega}\right]
\]

is just the interval $|q|<\frac{\sqrt{E}}{\omega}$; i.e., the major
axis of the energy ellipse. Given two points $Q,Q'\in M_{E}$ we have
the solution to the eikonal equation

\[
\sigma_{E}(Q,Q')=\int_{Q'}^{Q}\sqrt{2E-\omega^{2}q^{2}}dq
\]

Geometrically, this is area of the region bounded by the energy ellipse,
vertical axes at $Q,Q'$ and the horizontal axis. $\sigma_{E}(Q,Q')$
is a metric (in the sense of topology) on the above interval . The
maximum of $\sigma_{E}(Q,Q')$ occurs when $Q=-\frac{\sqrt{E}}{\omega},Q'=-\frac{\sqrt{E}}{\omega}$
and is equal to half the area of the energy ellipse. Thus,

\[
\sigma_{E}(Q,Q')\leq\pi\frac{E}{\omega}
\]

\textbf{}

This is reminiscent of Myer's inequality in Riemannian geometry\cite{doCarmoRiemGeom}.
If the Ricci tensor is bounded below 

\[
R_{ij}\xi^{i}\xi^{j}\geq\omega^{2}\xi^{i}\xi^{j}g_{ij},\quad\omega>0
\]

the distance between any two points in the manifold is bounded as
well:

\[
d(Q,Q')\leq\frac{\pi}{\omega}.
\]

Could there be a generalization of Myer's theorem to more general
mechanical systems with a convex, time-symmetric Hamiltonian?

\includegraphics{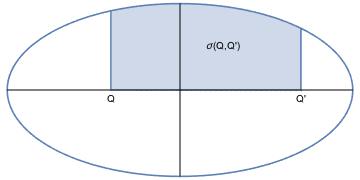}

The inverted harmonic oscillator 

\[
H=\frac{1}{2}\left[p^{2}-\omega^{2}q^{2}\right]
\]
has an unstable equilibrium point at $q=0=p$; it has constant negative
curvature 

\[
\mathcal{R}_{11}=-\omega^{2}
\]

and is the mechanical analogue of Lobachewski space.

\subsection{Constant Magnetic Field}

In the case of a particle moving on the plane we denote the co-ordinates
by $z\equiv(x,y)$ instead of $q^{1},q^{2}$. The Hamiltonian

\[
H=\frac{1}{2}\left[p_{x}+By\right]^{2}+\frac{1}{2}\left[p_{y}-Bx\right]^{2}
\]

corresponds to a particle in a constant magnetic field $B$ normal
to the plane. (We choose units where the mass and charge are equal
to one. So $B$ is just the cyclotron frequency.)

The curvature is constant and positive:

\[
\mathcal{R}_{ij}=\frac{1}{4}B^{2}\delta_{ij}.
\]

Hamilton's equations are equivalent to the Lorentz equations

\[
\ddot{x}=B\dot{y},\quad\ddot{y}=-B\dot{x}
\]

It is instructive to find the action $s_{T}(Z,Z')$ of the solution
satisfying the boundary conditions

\[
x(0)=X',\quad x(T)=X
\]

\[
y(0)=Y'\quad y(T)=Y
\]

This is a standard exercise in physics textbooks\cite{FeynmanHibbsMagField}
. After a long but straightforward calculation we get 

\[
s_{T}(Z,Z')=\left[\frac{1}{4}B\cot\frac{BT}{2}\right]\mid Z-Z'|^{2}-\frac{1}{2}BZ\times Z'
\]

where the cross-product is $Z\times Z'=XY'-YX'$.

We can verify directly that this satisfies the Hamilton-Jacobi equation

\[
\frac{1}{2}\left[\partial_{X}s_{T}+BY\right]^{2}+\frac{1}{2}\left[\partial_{Y}s_{T}-BX\right]^{2}+\frac{\partial s_{T}}{\partial T}=0.
\]

Its Legendre transform

\[
\sigma_{E}(Z,Z')=\min_{T}\left[ET+s_{T}(Z,Z')\right]
\]

\[
=\frac{2E}{B}\mathrm{arcsin}\left[\frac{B|Z'-Z|}{2\sqrt{2E}}\right]+\frac{1}{2}|Z'-Z|\sqrt{2E-\frac{1}{4}|Z'-Z|^{2}B^{2}}-\frac{1}{2}B\ Z\times Z
\]

Since the trajectory is a circle of radius $\frac{\sqrt{2E}}{B}$,
only points with $|Z-Z'|<2\frac{\sqrt{2E}}{B}$ are connected by a
smooth trajectory. Farther points would be connected by stitching
together piecewise-circular segments. The above formula describes
only one such segment.

Again we can verify directly that the stationary HJ equation is satisfied:

\[
\frac{1}{2}\left[\partial_{X}\sigma_{E}+BY\right]^{2}+\frac{1}{2}\left[\partial_{Y}\sigma_{E}-BX\right]^{2}=E
\]

In units where $E=\frac{1}{2}$ , (i.e., unit velocity) 

\[
\sigma(Z,Z')=\frac{1}{B}\mathrm{arcsin}\left[\frac{B|Z'-Z|}{2}\right]+\frac{1}{2}|Z'-Z|\sqrt{1-\frac{1}{4}|Z'-Z|^{2}B^{2}}-\frac{1}{2}B\ Z\times Z'
\]

We can understand each of these terms geometrically. The trajectory
is a circle of radius $\frac{\sqrt{2E}}{B}$ connecting $Z$ to $Z'$
. If \textbf{$B>0$} it is described in a counter-clockwise direction.
Let $C$ be the center and let $M$ be the point halfway on the chord
$ZZ'$ . Consider the right triangle $CMZ'$. The lengths of its sides
are 

\[
|CZ'|=\frac{1}{B},\quad|MZ'|=\frac{1}{2}|Z'-Z|,\quad|CM|=\frac{1}{B}\sqrt{1-\frac{1}{4}|Z'-Z|^{2}B^{2}}
\]

The half-angle at the center is 

\[
\widehat{MCZ'}=\mathrm{arcsin}\left[\frac{B|Z'-Z|}{2}\right]=\widehat{MCZ}
\]

Thus each term in $\sigma(Z,Z')$ has a meaning of an area (times
$B$), as illustrated in the figure:
\begin{itemize}
\item the first term is the area of the circular arc of angle $\widehat{MCZ}$
(Blue)
\item the second term is the area of the right triangle $CMZ'$(Yellow)
\item the third term subtracts the area of the triangle $OZZ'$ (Red)
\end{itemize}
\includegraphics{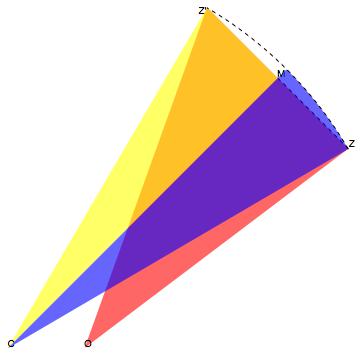}

\subsection{Magnetic Field Plus Quadratic potential}

We can combine the above two cases and add an extra dimension to get 

\[
H=\frac{1}{2}\left[p_{x}+By\right]^{2}+\frac{1}{2}\left[p_{y}-Bx\right]^{2}+\frac{1}{2}p_{z}^{2}+\phi
\]

where $\phi$ is a positive quadratic form in $x,y,z$. This describes
a particle in a Penning trap or the immediate vicinity of a Lagrange
point in the circular restricted three-body problem. (In the co-rotating
frame of the primary bodies, there is a Coriolis force which is mathematically
identical to the force due to a constant magnetic field normal to
the plane of rotation.) The curvature 

\[
\mathcal{R}_{ij}=\frac{1}{4}F_{ik}F_{jl}g^{kl}+\partial_{i}\partial_{j}\phi
\]

can be written conveniently in the co-ordinate system which diagonalizes
$\partial_{i}\partial_{j}\phi$

\[
\partial_{i}\partial_{j}\phi=\left(\begin{array}{ccc}
k_{1} & 0 & 0\\
0 & k_{2} & 0\\
0 & 0 & k_{3}
\end{array}\right)
\]

\[
F_{ij}=\left(\begin{array}{ccc}
0 & B & 0\\
-B & 0 & 0\\
0 & 0 & 0
\end{array}\right)
\]

\[
\mathcal{R}_{ij}=\left(\begin{array}{ccc}
k_{1}+\frac{1}{4}B^{2} & 0 & 0\\
0 & k_{2}+\frac{1}{4}B^{2} & 0\\
0 & 0 & k_{3}
\end{array}\right)
\]

If $\phi$ is harmonic (e.g., an electrostatic field as in the Penning
trap or a Newtonian Gravitational field as in the three-body problem) 

\[
k_{1}+k_{2}+k_{3}=0.
\]

It is well known that such a harmonic potential $\phi$ does not have
a stable equilibrium as at least one of the $k_{i}$ must be negative.
Adding a strong enough magnetic field can stabilize such a potential;
this is the idea behind the Penning trap and the surprising stability
of the Lagrange points $L_{4}$ and $L_{5}$. 

Whether harmonic or not, the case 

\[
k_{3}>0,\quad k_{1}<0,\quad k_{2}<0,\quad\frac{1}{2}B^{2}>\sqrt{k_{1}k_{2}}+\frac{|k_{1}|+|k_{2}|}{2}
\]

is known to be stable\cite{RajeevMechanics}. 

In this case, if the curvature is positive,

\[
k_{3}>0,\quad\frac{1}{4}B^{2}>|k_{1}|,\quad\frac{1}{4}B^{2}>|k_{2}|
\]

it follows that $\frac{1}{4}B^{2}$ is also greater than the average
of the geometric and arithmetic means of the r.h.s.:

\[
\frac{1}{4}B^{2}>\frac{1}{2}\left[\sqrt{k_{1}k_{2}}+\frac{|k_{1}|+|k_{2}|}{2}\right].
\]

Thus positivity of curvature is sufficient for stability in this case.
It is not necessary: we can have 

\[
|k_{2}|<\frac{1}{4}B^{2}<|k_{1}|
\]

and still have 

\[
\frac{|k_{1}|+|k_{2}|}{2}<\frac{1}{4}B^{2}.
\]

Since the arithmetic mean of positive numbers always exceed their
geometric mean,

\[
\sqrt{k_{1}k_{2}}<\frac{|k_{1}|+|k_{2}|}{2}
\]

this would give stability without positivity of curvature.

On the other hand, negative curvature is sufficient for instability:
$k_{3}<0$.

\section{Acknowledgement}

I thank Sumit Das for explaining that the scalar field $\phi$ is
the dilaton. In addition thanks to Miguel Alonso, Alex Iosevich, Andrew
Jordan, Arnab Kar, Govind Krishnaswami and Evan Ranken for discussions.


\begin{thebibliography}{10}
\bibitem{doCarmoRiemGeom}M. P. do Carmo, \emph{Riemannian Geometry,
}Birkhauser (1992).

\bibitem{ChernFinsler}S. S. Chern, \emph{Geometry without the Quadratic
Restriction, Notices of the AMS, }959 (1996); D. Bao, R. L. Bryant,
S. S. Chern and Z. Shen, \emph{A Sampler of Riemann\textendash Finsler
Geometry}, Cambridge University Press (2004).

\bibitem{RandersGRUnifiedEMPhysRev.59.195}G. Randers, Phys. Rev.
\textbf{59,} 195 (1941).

\bibitem{Hamilton}W. R. Hamilton, Trans. Roy. Irish Acad., 17, 1\textendash 144
(1837).

\bibitem{Klimes}L. Klimes, Journal of Electromagnetic Waves and Applications,
\textbf{27,}1589(2013).

\bibitem{WeinsteinLagrSubMflds}A. Weinstein, Adv. Math. \textbf{6,
}329 (1971).

\bibitem{BambusiBirkhoffNormalForm}D. Bambusi \textquotedbl{}Birkhoff
normal form and almost global existence for some Hamiltonian PDEs.\textquotedbl{}
(2007). Available at http://users.mat.unimi.it/users/bambusi/pedagogical.pdf 

\bibitem{HendersonRajeev} R.J. Henderson and S.G. Rajeev, Class.Quant.Grav.
\textbf{11}, 1631 (1994), arXiv:gr-qc/9401029.

\bibitem{SymplecticConnections}P.Bieliavsky, M.Cahen, S. Gutt and
J. Rawnsley J. Geom. Phys. \textbf{38,}140 (2001\textbf{)};P.Bieliavsky,
M.Cahen, S. Gutt, J. Rawnsley and L. Schwachhofer, \emph{Symplectic
Connections}, arXiv:math/0511194 {[}math.SG{]}; K. Habermann and L.
Habermann, \emph{Introduction to Symplectic Dirac Operators, }Springer
(2006)

\bibitem{ArnoldCurvature} V. I. Arnold, Ann. Inst. Poly. Genoble
\textbf{16 }, 319 (1966) 

\bibitem{Govind3Body}G. S. Krishnaswami and H. Senapati J. Math.
Phys. \textbf{57}, 102901 (2016), arXiv:1606.05091.

\bibitem{MontgomerySubRiemGeom}R. Montgomery, \emph{A Tour of Subriemannian
Geometries, Their Geodesics and Applications,} AMS (2002)

\bibitem{JohnLeeRiemannianManifolds}J. M. Lee, \emph{Riemannian Manifolds}
Springer (1997)

\bibitem{FeynmanHibbsMagField} Problem 3-10 in R. P. Feynman and
A. R. Hibbs, \emph{Quantum Mechanics and Path Integrals, }McGraw-Hill
(1965)

\bibitem{RajeevMechanics}S. G. Rajeev, \emph{Advanced Mechanics},
Oxford (2012).
\end{thebibliography}
\end{document}